%% file: manu.tex
\documentclass[aps,prx,twocolumn,superscriptaddress,a4paper,english,longbibliography]{revtex4-1}
\usepackage{times}
\usepackage{color}
\usepackage[colorlinks=true,urlcolor=blue,citecolor=blue]{hyperref}
\usepackage{url}
\usepackage{breakurl}
\usepackage{soul}
\usepackage{graphicx}
\usepackage{amsmath}
\usepackage{subfigure}
\usepackage{xcolor}
\usepackage{amssymb}
\usepackage{latexsym}
\usepackage{hyperref}% add hypertext capabilities
\DeclareGraphicsExtensions{.png,.eps}
\usepackage{float}
\usepackage{appendix}
\usepackage{etoolbox}
\usepackage{mathtools}

\usepackage{xcolor}
\usepackage[urlcolor=blue]{hyperref}      % links to citations
\hypersetup{
	colorlinks = true,                    % text and not border
	citecolor = {blue},
	linkcolor = {purple},
}

\begin{document}	
	
\title{Functional Tensor Network Solving Many-body Schr\"odinger Equation}

\author{Rui Hong}
\affiliation{Department of Physics, Capital Normal University, Beijing 100048, China}
\author{Ya-Xuan Xiao}
\affiliation{Department of Physics, Capital Normal University, Beijing 100048, China}
\author{Jie Hu}
\affiliation{Department of Physics, Capital Normal University, Beijing 100048, China}
\author{An-Chun Ji}
\affiliation{Department of Physics, Capital Normal University, Beijing 100048, China}
\author{Shi-Ju Ran}\email[Corresponding author. Email: ]{sjran@cnu.edu.cn}
\affiliation{Department of Physics, Capital Normal University, Beijing 100048, China}

\date{\today}
\begin{abstract}
	Schr\"odinger equation belongs to the most fundamental differential equations in quantum physics. However, the exact solutions are extremely rare, and many analytical methods are applicable only to the cases with small perturbations or weak correlations. Solving the many-body Schr\"odinger equation in the continuous spaces with the presence of strong correlations is an extremely important and challenging issue. In this work, we propose the functional tensor network (FTN) approach to solve the many-body Schr\"odinger equation. Provided the orthonormal functional bases, we represent the coefficients of the many-body wave-function as tensor network. The observables, such as energy, can be calculated simply by tensor contractions. Simulating the ground state becomes solving a minimization problem defined by the tensor network. An efficient gradient-decent algorithm based on the automatically differentiable tensors is proposed. We here take matrix product state (MPS) as an example, whose complexity scales only linearly with the system size. We apply our approach to solve the ground state of coupled harmonic oscillators, and achieve high accuracy by comparing with the exact solutions. Reliable results are also given with the presence of three-body interactions, where the system cannot be decoupled to isolated oscillators. Our approach is simple and with well-controlled error, superior to the highly-nonlinear neural-network solvers. Our work extends the applications of tensor network from quantum lattice models to the systems in the continuous space. FTN can be used as a general solver of the differential equations with many variables. The MPS exemplified here can be generalized to, e.g., the fermionic tensor networks, to solve the electronic Schr\"odinger equation.
\end{abstract}
\maketitle

\section{Introduction}

Solving differential equations belongs to the most fundamental but challenging tasks in mathematics, physics, and etc. In general, the situations where we have the exact solutions are extremely rare, thus various analytical and numerical methods were developed under the simplifications or approximations of different extent.

Let us concentrate on quantum physics, where the Schr\"odinger equation plays a fundamental role. Different approximative treatments made to this equation have evolved into different sub-fields. For instance, the density functional theories and the so-called \textit{ab-initio} calculations (see, e.g.,~\cite{GDL03DFT, CU12abinitioBook}) successfully predict the properties of countless quantum matters ranging from molecules to solids, under the assumption of weak correlations. Recently, the hybridization with machine learning has triggered a new upsurge in studying the Schr\"odinger equation, including the simulations of the ground states by better considering correlations~\cite{schutt_unifying_2019, han_solving_2019, deep-neural2020, manzhos_machine_2020, CMC20fermionicNN} and inversely predicting the potentials knowing the wave-functions or relevant physical information~\cite{sehanobish_learning_2021, hong_predicting_2021, VKLJK21MLpotential, WFC21MLpotential}.

Towards the strongly-correlated cases, an important direction is simplifying to the quantum lattice models, such as Heisenberg or Hubbard models on discretized lattices. Among the successful algorithms, remarkable progresses have been made based on tensor network (TN) ~\cite{VMC08MPSPEPSRev, CV09TNSRev, O14TNSRev, RTPC+17TNrev, O19TNrev}. As two important examples, we have the density matrix renormalization group for simulating the ground states of one- and quasi-one-dimensional systems~\cite{W92DMRG,W93DMRG,S05DMRGrev,S11DMRGRev, SW12DMRG2DRev}, and projected-entangled pair states for the higher-dimensional ones~\cite{VWPC06PEPSfamous, JOVVC08PEPS}.

The success of TN lies in its high efficiency of representing quantum many-body states as well as the powerful algorithms to deal with the TN calculations. On the cost of obeying the area law of entanglement entropy~\cite{VWPC06PEPSfamous, ECP10AreaLawRev, GE16arealawTNS}, TN reduces the exponential complexity of representing a many-body state to be polynomial. Accurate results are obtained by TN, thanks to the fact that for most models we are interested in, such as those with local interactions in one dimension~\cite{VC06MPSFaithfully}, the area law holds. However, beyond the quantum lattice models, TN for those with continuous variables are mainly concentrated on the quantum fields~ \cite{VC10cMPS, HCOV13cMPS, SRHE14fieldcMPS, JBHOV15cTNS, TC19cTNS, AHCS21fieldTNS}. The use of TN for solving differential equations with many continuous variables is unexplored. 

\begin{figure}[tbp]
	\centering
	\includegraphics[angle=0,width=1\linewidth]{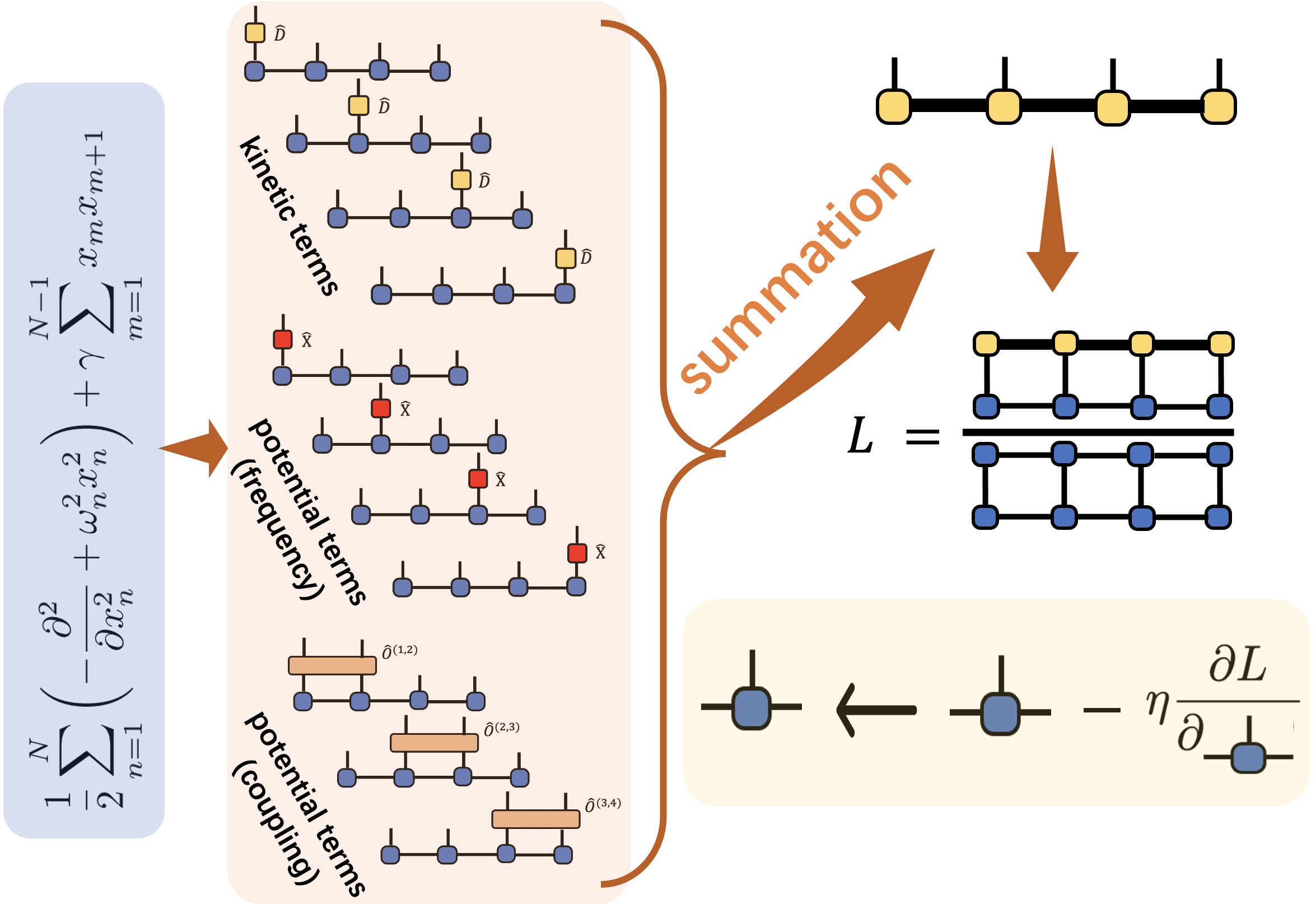}
	\caption{(Color online) The illustration of the functional MPS approach. By representing the trial wave-function in MPS, the loss function $L$ becomes the inner product of two MPS's. One MPS is the summation of several MPS's, of which each is resulted from the trial wave-function acted by an operator. The inset illustrates the gradient descent [Eq. (\ref{eq-gradMPS})] to update the tensors in the trial MPS.}
	\label{fig-idea}
\end{figure}

In this work, TN is proposed to solve the many-body Schr\"odinger equation in the continuous space. Given the orthonormal functional bases, the coefficients of the quantum wave-function is represented in the form of TN. Defining the loss function $L$ as the energy, the automatically differentiation technique~\cite{LLWX10diffTN} is utilized to achieve the TN representing the ground state. We dub this approach as \textit{functional TN}. Taking the matrix product state (MPS in short, which is a special one-dimensional TN)~\cite{VMC08MPSPEPSRev} as an example, our approach is illustrated in Fig. \ref{fig-idea}. The loss function is calculated as the inner product of two MPS's. One MPS is the summation of many MPS's representing the wave-function acted by the corresponding operators, and the other is the MPS representing the wave-function itself. The tensors forming the MPS are automatically differentiable, and are updated by the gradient descent algorithm. The gradients are obtained in the back propagation process~\cite{LLWX10diffTN}, similar to the optimizations of neural networks. 

We test our approach on the coupled harmonic oscillators with two- and three-body interactions [see Eq. (\ref{eq-HOH})]. The model cannot be decoupled to isolated oscillators with the presence of the three-body terms. With only the two-body interactions, high accuracy is demonstrated by comparing the achieved ground-state energy to the exact one. The error is well controlled by the entanglement of the MPS. The ground-state energy and entanglement entropy with the three-body terms are also demonstrated. Compared with the solvers of differential equations based on neural networks \cite{MTJ95NNDE, LLF98NNODE, MC16MLODE, BHHB19MLODE, SHYZ+19NNPDE} that are in general highly non-linear, our functional TN solver does not require sampling or the training data, thus do not belong to the ``data-driven'' solvers. The optimizations of TN are implemented simply by tensor contractions. Our works sheds light on using TN as an efficient solver of the general many-variable differential equations in and beyond quantum physics.

\section{Preliminaries and notations}
\subsection{Basis and expansion}

With a set of orthonormal functional basis $\{\phi_{s}(x)\}$ satisfying $\int_{-\infty}^{\infty} \phi_{s'}^*(x)\phi_s(x)dx=\delta_{s's}$, a given function $\psi(x)$ can be expanded as
 \begin{eqnarray}
	\psi(x)=\sum_{s=0}^{\mathcal{D}-1}C_s\phi_s(x),
	\label{eq-expand}
\end{eqnarray}
where $C_s$ denotes the expansion coefficients and $\mathcal{D}$ is the expansion order. Here, we assume $\psi(x)$ to be smooth. 

Considering $\psi(x)$ to be the wave-function of a quantum system, it should satisfy the normalization condition as $\int_{-\infty}^{\infty} \psi^{\ast}(x)\psi(x) dx = 1$. Thanks to the orthonormal condition of the basis, the normalization condition can be represented by the L2-norm of the coefficients as 
 \begin{eqnarray}
	|\mathbf{C}| = \sqrt{\sum_{s} |C_s|^{2}} = 1.
	\label{eq-normalization}
\end{eqnarray}
Note we use a bold letter to denote a vector, matrix or tensor, such as $\mathbf{C}$, and use the same letter with lower indexes to denote its elements, such as the $s$-th element $C_s$.

\subsection{Operations}

Consider an operator, denoted as $\hat{O}$, that satisfies the linear condition in the functional space as
 \begin{eqnarray}
	\hat{O}\psi(x)=\hat{O}\sum_sC_s\phi_s(x)=\sum_sC_s\hat{O}[\phi_s(x)].
	\label{eq-operate}
\end{eqnarray}
Assume for each basis function $\phi_s(x)$, $\hat{O}$ satisfies 
 \begin{eqnarray}
	\hat{O}[\phi_s(x)]=\sum_{s'=0}^{\mathcal{D}-1} O_{s's}\phi_{s'}(x).
	\label{eq-operate_basis}
\end{eqnarray}
Apparently, the expansion coefficients of the function $\tilde{\psi}(x)=\hat{O}[\psi(x)]=\sum_s\tilde{C}_s\phi_s(x)$ satisfy
\begin{eqnarray}
	\tilde{C}_{s'}=\sum_{s}O_{s's}C_{s}.
	\label{eq-op_matrix}
\end{eqnarray}
In general, $O_{s's}$ can be numerically evaluated as
\begin{eqnarray}
	O_{s's}=\int_{-\infty}^{\infty} \phi^*_{s'}(x)\hat{O}[\phi_s(x)]dx.
	\label{eq-op_int}
\end{eqnarray}

In some special cases, the matrix $O_{s's}$ given the basis $\{\phi_s(x)\}$ can be solved analytically. As an example, let us take $\phi_s(x)$ as the $s$-th eigenstate of the quantum harmonic oscillator with the Hamiltonian
\begin{eqnarray}
	\hat{H}^{\text{HO}} = -\frac{1}{2} \frac{d^{2}}{dx^{2}}+ \frac{x^{2}}{2}.
	\label{eq-hamilt_1dHO}
\end{eqnarray}
We take the Plank constant $\hbar=1$ for simplicity. We have
\begin{eqnarray}
	\phi_s(x)=\left(\frac{1}{2^ss!\sqrt{\pi}}\right)^{\frac{1}{2}}e^{-\frac{x^2}{2}}h_s(x),
	\label{eq-HOphi}
\end{eqnarray}
with $h_s(x)$ the Hermitian polynomial. We dub Eq. (\ref{eq-HOphi}) as single-oscillator basis (SOB). Obviously $\{\phi_s(x)\}$ satisfy the orthonormal conditions. 

We now consider the operation $\hat{D} = \frac{d}{dx}$. With $D_{s's} = \int_{-\infty}^{\infty} \phi^*_{s'}(x)\hat{D}[\phi_s(x)]dx$ [Eq. (\ref{eq-op_int})], we have 
\begin{equation}
\label{eq-D}
D_{s's}=\left\{
		\begin{array}{lr}
		\sqrt{\frac{s+1}{2}}, & s'=s-1; \\
		-\sqrt{\frac{s+1}{2}}, & s'=s+1. \\
		\end{array}
\right.
\end{equation}
The dimensions of the matrix $\mathbf{D}$ should be infinite (i.e., $\mathcal{D} \to \infty$) to exactly represent the differential operator. In practice, one may use a proper approximation by taking a finite $\mathcal{D}$. For the $k$-order differentiation, we have
\begin{eqnarray}
	\int_{-\infty}^{\infty} \phi^*_{s'}(x) \hat{D}^{k} [\phi_s(x)]dx = [\mathbf{D}^{k}]_{s's},
	\label{eq-Dk}
\end{eqnarray}
with $\mathbf{D}^{k}$ the $k$-th power of the matrix $\mathbf{D}$. Another example is the operation $\hat{X} = x$. Similarly, we have 
\begin{equation}
\label{eq-X}
X_{s's}=\left\{
		\begin{array}{lr}
		\sqrt{\frac{s+1}{2}}, & s'=s-1; \\
		\sqrt{\frac{s+1}{2}}, & s'=s+1. \\
		\end{array}
\right.
\end{equation}

\subsection{Solving differential equation by optimization}
\label{sec-tensor}

Consider a differential equation formed by $P$ terms. We can formerly write it as
\begin{eqnarray}
	\sum_{p=1}^{P} \hat{O}^{[p]}[\psi(x)]=0
	\label{eq-DE}
\end{eqnarray}
Let us take the the static Schr\"odinger equation as an example, we have
\begin{eqnarray}
\left\{
		\begin{array}{lr}
		\hat{O}^{[1]}=-\frac{1}{2}\frac{d^2}{dx^2},\\
		\hat{O}^{[2]}=V(x), \\
		\hat{O}^{[3]}=-E. \\
		\end{array}
\right.
\end{eqnarray}
The first two terms (kinetic and potential terms, respectively) correspond to the Hamiltonian $\hat{H} = \hat{O}^{[1]} + \hat{O}^{[2]}$, and the third corresponds to the (negative) energy.

To solve the static Schr\"odinger equation, we regard the coefficients $C_s$ in $\psi(x) = \sum_s C_s \phi_s(x)$ as the variational parameters. Now we consider the calculation of the ground state, i.e., the eigen-function with the lowest eigenvalue (energy) of the given Hamiltonian. With the normalization constraint of the wave-function given in Eq. (\ref{eq-normalization}), the energy (quantum average of $\hat{H}$) satisfies 
\begin{equation}
	E = \langle \hat{H} \rangle = \int_{-\infty}^{\infty} \psi^{\ast}(x) \hat{H} \psi(x) dx = \sum_{s's} C^{\ast}_{s} H_{ss'} C_{s'},
	\label{eq-energy}
\end{equation}
where $H_{mn}$ can be calculated using Eq. (\ref{eq-op_int}). Define the loss function 
\begin{eqnarray}
	L = \frac{E}{|\mathbf{C}|^{2}}.
	\label{eq-loss}
\end{eqnarray}
We introduce the devision over $|\mathbf{C}|^{2}$ so that we do not need to consider the normalization of $\psi(x)$ in the optimization process. Then $\mathbf{C}$ is iteratively updated using gradient descent as
\begin{eqnarray}
	\mathbf{C} \leftarrow \mathbf{C} - \eta \frac{\partial L}{\partial \mathbf{C}}.
	\label{eq-grad}
\end{eqnarray}
with $\eta$ the learning rate or gradient step.

Beyond the static Schr\"odinger equation, we can use the gradient descent to solve a differential equation given in Eq. (\ref{eq-DE}). In a given set of functional basis, the differential equation can be written as,
%$\sum_ia_i\hat{O}_i\psi(x)=0$
\begin{eqnarray}
\sum_{ps's}O_{s's}^{[p]}C_s\phi_{s'}(x)=0
\end{eqnarray}
Since $\{\phi_s(x)\}$ is a set of orthonormal basis, we have
\begin{eqnarray}
\sum_{ps}O_{s's}^{[p]}C_s=0, \text{ for }{\forall} s'.
\end{eqnarray}
 Introduce the vector $\mathbf{Z}$ with its $s'$-th element $Z_{s'}=\sum_{ps}O_{s's}^{[p]}C_s$. With $|\mathbf{Z}|^2=0$, $\psi^{\ast}(x)$ is the solution of the differential equation. Therefore, we define the loss function as,
\begin{eqnarray}
\mathcal{L}=|\mathbf{Z}|^2.
\label{eq-mathcalL}
\end{eqnarray}
The coefficients can be updated using Eq. (\ref{eq-grad}).

\section{Functional matrix product state}

\subsection{Matrix product state representation for the coefficients of multi-variable function}

Given $N$ sets of orthonormal bases $\{\phi_{s_n}(x_n)\}$, a function with $N$ independent variables $\mathbf{x} = (x_1,\cdots, x_N)$ can be expanded as 
 \begin{equation}
	\psi(\mathbf{x}) = \sum_{s_1\cdots s_N=0}^{\mathcal{D}-1} C_{s_1\cdots s_N} \phi_{s_1}(x_1)\cdots\phi_{s_N}(x_N).
	\label{eq-mfun}
\end{equation}
Obviously, the complexity of the coefficient tensor $\textbf{C}$ scales exponentially with the number of variables $N$ as $O(\mathcal{D}^N)$. 

One key of our proposal is using TN to represent the coefficients. As illustrated in Fig.~\ref{fig-op}(a), we take MPS as an example and have
 \begin{equation}
	C_{s_1\cdots s_N}=\sum_{\alpha_0\cdots \alpha_{N}=0}^{\chi-1} A_{\alpha_0 s_1\alpha_1}^{(1)}A_{\alpha_1s_2\alpha_2}^{(2)}\cdots A_{\alpha_{N-1}s_{N} \alpha_N}^{(N)},
	\label{eq-MPS}
\end{equation}
with $\{\alpha_{n}\}$ the virtual bonds. We take the MPS to have the open boundary condition in the whole paper, with $\dim(\alpha_0) = \dim(\alpha_N) = 1$. The upper bound of $\dim(\alpha_{n})$ ($n=1, \cdots, N-1$) is called the virtual bond dimension of the MPS, denoted by $\chi$. The indexes $\{s_{n}\}$ are called the physical bonds, and the dimension is called the physical bond dimension. In our cases, we have $\dim(s_{n})= \mathcal{D}$, i.e., the expansion order. The number of parameters in the MPS (i.e., the total number of elements in the tensors $\{\mathbf{A}^{(n)}\}$ for $n=1, \cdots, N$) scales only linearly with $N$ as $O(N\mathcal{D}\chi^{2})$, while that of $\mathbf{C}$ scales exponentially as $O(\mathcal{D}^{N})$. 

Akin to the one-variable cases, the norm of $\psi(\mathbf{x})$ equals to the norm of the coefficient tensor (or MPS), i.e.,
\begin{equation}
\int_{-\infty}^{\infty} \psi^{\ast}(\mathbf{x}) \psi(\mathbf{x}) d\mathbf{x} = \sum_{s_1,\cdots, s_N} C^{\ast}_{s_1,\cdots, s_N} C_{s_1,\cdots, s_N} = |\mathbf{C}|^{2},
\label{eq-Z2}
\end{equation}
with $d\mathbf{x} = \prod_{n=1}^{N} dx_{n}$. Note that when the tensors in the MPS $\{\mathbf{A}^{(n)}\}$ are given, $|\mathbf{C}|^{2}$ can be obtained without calculating $\textbf{C}$. Therefore, the exponential complexity is avoided. Notably, the proposed method and discussions in this work can be readily extended to generally other TN's.

\subsection{Operations and quantum average}
\label{sec-mps-op}

Consider an operation $\hat{O}^{(m)}$ on $x_{m}$. According to the linearity and the independency of the variables, we have
\begin{eqnarray}
\tilde{\psi}(\mathbf{x}) &=& \hat{O}^{(m)} [\psi(\mathbf{x})] \nonumber \\ 
&=& \sum_{s_1\cdots s_N=0}^{\mathcal{D}-1} C_{s_1,\cdots, s_N} [\prod_{n \neq m} \phi_{s_n}(x_n)] \hat{O}^{(m)}[\phi_{s_m}(x_m)].\nonumber \\ 
\end{eqnarray}
Denote the tensors in the MPS representing $\tilde{\psi}(\mathbf{x})$ as $\{\mathbf{\tilde{A}}^{(n)}\}$. Given the tensors $\{\mathbf{A}^{(n)}\}$ in the MPS representation of $\psi(\mathbf{x})$, we have $\mathbf{A}^{(n)} = \mathbf{\tilde{A}}^{(n)}$ for $n \neq m$. For $n = m$, we have
\begin{eqnarray}
\tilde{A}^{(m)}_{\alpha s \alpha'} = \sum_{s'} O^{(m)}_{ss'} A^{(m)}_{\alpha s' \alpha'},
\label{eq-MPStensorO}
\end{eqnarray}
where $O^{(m)}_{ss'}$ satisfies Eq. (\ref{eq-op_int}). See Fig.~\ref{fig-op}(b) for an illustration.

Similarly, the MPS obtained by acting multiple operators can be derived. Take two operators $\hat{O}^{(m_1)}$ and $\hat{O}^{(m_2)}$ as an example. The tensors $\{\mathbf{\tilde{A}}^{(n)}\}$ in the MPS representing $\tilde{\psi}(\mathbf{x}) = \hat{O}^{(m_2)} \hat{O}^{(m_1)} \psi(\mathbf{x})$ satisfy $\mathbf{\tilde{A}}^{(n)} = \mathbf{A}^{(n)}$ for $n \neq m_1$ and $n \neq m_2$, and $\tilde{A}^{(n)}_{\alpha s \alpha'} = \sum_{s'} O^{(n)}_{ss'} A^{(n)}_{\alpha s' \alpha'}$ for $n = m_1$ or $n = m_2$.

Consider an operator acting on multiple variables, such as $\hat{O}^{(m_1, m_2)}$ acting on $x_{m_1}$ and $x_{m_2}$. We assume that $\hat{O}^{(m_1, m_2)}$ cannot be decomposed to the product of two single-variable operators, i.e., $\hat{O}^{(m_1)} \hat{O}^{(m_2)}$. Then as an extension of Eq. (\ref{eq-op_int}), we introduce a fourth-order tensor $\mathbf{O}^{(m_1, m_2)}$, where its elements satisfy
\begin{eqnarray}
	O^{(m_1, m_2)}_{n_1'n'_2n_1n_2} && =  \int_{-\infty}^{\infty} \phi^*_{n'_1}(x_{m_1}) \phi^*_{n'_2}(x_{m_2}) \nonumber \\ 
	&&\hat{O}^{(m_1, m_2)}[\phi_{n_1}(x_{m_1}) \phi_{n_2}(x_{m_2})] dx_{m_1} dx_{m_2}.
	\label{eq-multi_op}
\end{eqnarray}
The calculation of the MPS representing $\tilde{\psi}(\mathbf{x}) = \hat{O}^{(m_1, m_2)}[\psi(\mathbf{x})]$ is illustrated in Fig.~\ref{fig-op}(c). The actions of multi-variable operators can be similarly defined.

Consider the quantum average of the operator $\hat{O}^{(m)}$ as
\begin{eqnarray}
	\langle \hat{O} \rangle = \int_{-\infty}^{\infty} \psi^{\ast}(\mathbf{x}) \hat{O}^{(m)}[\psi(\mathbf{x})] d\mathbf{x}.
\end{eqnarray}
As illustrated in Fig.~\ref{fig-op}(d), $\langle \hat{O} \rangle$ is calculated in the same way as calculating the average of a single-site operator with a standard MPS, similar to Eq. (\ref{eq-energy}). Same arguments can be made for the quantum average of multi-variable operators.

\subsection{Solving coupled harmonic oscillators}

\begin{figure}[tbp]
	\centering
	\includegraphics[angle=0,width=1\linewidth]{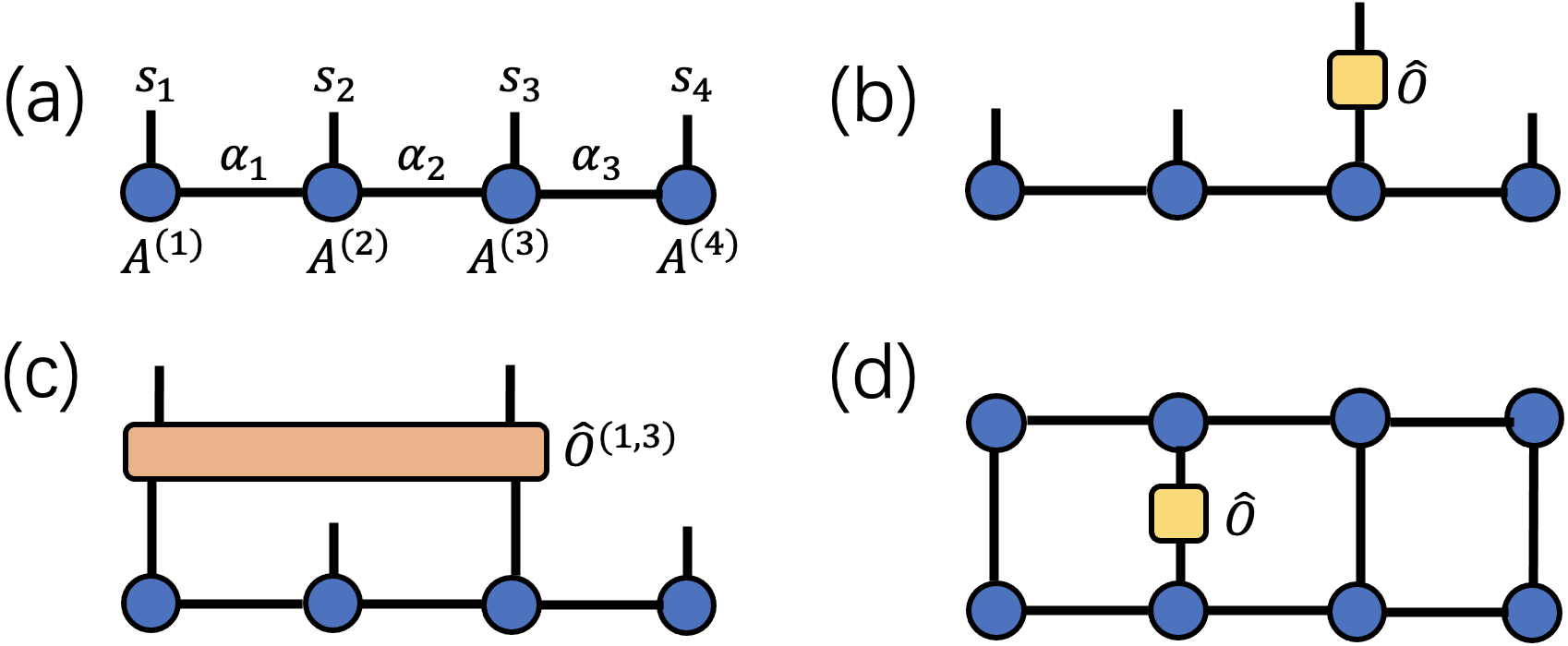}
	\caption{(Color online) (a) The graph of an MPS [Eq. (\ref{eq-MPS})] with its physical and virtual bonds. The virtual bonds at the two ends are one-dimensional, and thus are ignored in the graph. In (b) and (c), we illustrate the actions of one- and two-body operators on the MPS. In (d) we illustrate the average of an one-body operator $\langle \hat{O} \rangle$.}
	\label{fig-op}
\end{figure}

We consider the following $N$ coupled harmonic oscillators in one dimension as an example, where the Hamiltonian reads
\begin{eqnarray}
\hat{H}^{\text{HO}} =&& \frac{1}{2} \sum_{n=1}^{N} \left( -\frac{\partial^{2}}{\partial x_{n}^{2}} + \omega_n^2 x_n^2 \right) + \gamma \sum_{m=1}^{N-1} x_m x_{m+1} \nonumber
\\&& + \tilde{\gamma} \sum_{m=1}^{N-2} x_m x_{m+1} x_{m+2},
\label{eq-HOH}
\end{eqnarray}
where $\omega_n$ gives the natural frequency of the $n$-th oscillator, and $\gamma$ and $\tilde{\gamma}$ are the two- and three-body coupling constants, respectively.

We choose the bases $\{\phi_{s_n}(x_n)\}$ as SOB [Eq. (\ref{eq-HOphi})], considering that the matrices of the required operators, namely $\hat{D}$ and $\hat{X}$, can be analytically obtained [Eqs. (\ref{eq-D}) and (\ref{eq-X})]. Suppose the ground state $\psi(\mathbf{x})$ we aim to obtain is written in the MPS formed by the tensors $\{\mathbf{A}^{(n)}\}$. For the kinetic terms, we define $\tilde{\psi}^{\text{K}(m)} (\mathbf{x}) = -\frac{1}{2} \frac{\partial^{2}}{\partial x_{m}^{2}} \psi(\mathbf{x})$. According to Sec. \ref{sec-mps-op}, the tensors of the MPS representing $\tilde{\psi}^{\text{K}(m)}$ can be obtained, where the $m$-th tensor should be changed to
\begin{eqnarray}
  \tilde{A}^{(m)}_{\alpha s \alpha'} =-\frac{1}{2} \sum_{s's''} D_{ss'}  D_{s's''} A^{(m)}_{\alpha s'' \alpha'},
\end{eqnarray}
with $\mathbf{D}$ the coefficient matrix of the differential operator $\hat{D}$ in the SOB [Eq. (\ref{eq-D})]. Note the coefficients of operators only depend on the choice of basis [Eq. (\ref{eq-op_int})], instead of the number of variables or the form representing the coefficients of the wave-functions. 

For the potential terms, we define $\tilde{\psi}^{\text{P}(m)}(\mathbf{x}) = \frac{1}{2} \omega_{n}^{2} x_{m}^{2} \psi(\mathbf{x})$. Similarly, the MPS representation of $\tilde{\psi}^{\text{P}(m)}(\mathbf{x})$ can be obtained from $\{\mathbf{A}^{(n)}\}$ and $\mathbf{X}$ by using Eq. (\ref{eq-X}). For the coupling terms, we define $\tilde{\psi}^{\text{C}(m, m+1)}(\mathbf{x}) = \gamma x_{m} x_{m+1} \psi(\mathbf{x})$. The $m$- and $(m+1)$-th tensors should be calculated following Eq. (\ref{eq-MPStensorO}). The MPS's corresponding to the three-body interactions can be similarly defined.

In all, we have $(4N-3)$ MPS's, in which $N$ MPS's are from the kinetic terms, $N$ from the frequency terms, and $(2N-3)$ from the coupling terms. The summation of these MPS's results in the MPS  that represents $\tilde{\psi}^{\text{H}}(\mathbf{x}) \coloneqq \hat{H} \psi(\mathbf{x})$.
Two MPS's with the same physical bond dimension can be added, which results in an MPS with the same physical bond dimension. Therefore, we can obtain $\tilde{\psi}^{\text{H}}(\mathbf{x})$ as an MPS. Denoting the virtual bond dimensions of two added MPS's as $\chi_{1}$ and $\chi_{2}$, respectively, the virtual bond dimension of the resulting MPS satisfies $\chi \leq \chi_{1} + \chi_{2}$. The virtual bond dimension (denoted as $\chi_{\text{H}}$) of $\tilde{\psi}^{\text{H}}(\mathbf{x}) $ satisfies $\chi_{\text{H}} \leq (4N-3)
\chi$ with $\chi$ the virtual bond dimension of $\psi(\mathbf{x})$. Since the MPS's in the additions have many shared tensors, we in fact have $\chi_{\text{H}} \ll (4N-3)
\chi$. See more details in Appendix \ref{append-add}.

To obtain the ground state, we choose the energy in Eq. (\ref{eq-loss}) as the loss function $L$. With a trial MPS (where the tensors can be initialized randomly), $L$ can be calculated with polynomial complexity, avoiding the exponentially-large full coefficient tensor. For instance, the energy [Eq. (\ref{eq-energy})] is obtained by the inner product of the MPS's $\psi(\mathbf{x})$ and $\tilde{\psi}^{\text{H}}(\mathbf{x})$. The illustration of the inner product is similar to Fig. \ref{fig-op}(d). The complexity of calculating the inner product of two MPS's generally scales as $O[N\mathcal{D}\chi\chi_{H}(\chi + \chi_{H})]$. See more details in Appendix \ref{append-inner}. 

After calculating the loss $L$, the tensors in the MPS representing the wave-function $\psi(\mathbf{x})$ can be updated by the gradient descent as
\begin{eqnarray}
	\mathbf{A}^{(n)} \leftarrow \mathbf{A}^{(n)} - \eta \frac{\partial L}{\partial \mathbf{A}^{(n)}},
	\label{eq-gradMPS}
\end{eqnarray}
where the gradients $\frac{\partial L}{\partial \mathbf{A}^{(n)}}$ can be obtained by the automatic differentiation technique of TN~\cite{LLWX10diffTN}. In practice, we obtained the gradients for all tensors and update them simultaneously in a back-propagation process. We choose the Adam optimizer~\cite{KB15Adam} to control the learning rate $\eta$. After sufficiently many iterations of updates, $L$ converges to the ground-state energy, and $\psi(\mathbf{x})$ converges to the ground state. Compared with the solvers of differential equations based on neural networks that are highly non-linear~\cite{MTJ95NNDE, LLF98NNODE, MC16MLODE, BHHB19MLODE, SHYZ+19NNPDE}, one advantage of our functional MPS (and generally TN) solver is that sampling is not required. The optimization is implemented simply by tensor contractions.

\section{Numerical results}

\begin{figure}[tbp]
	\centering
	\includegraphics[angle=0,width=0.9\linewidth]{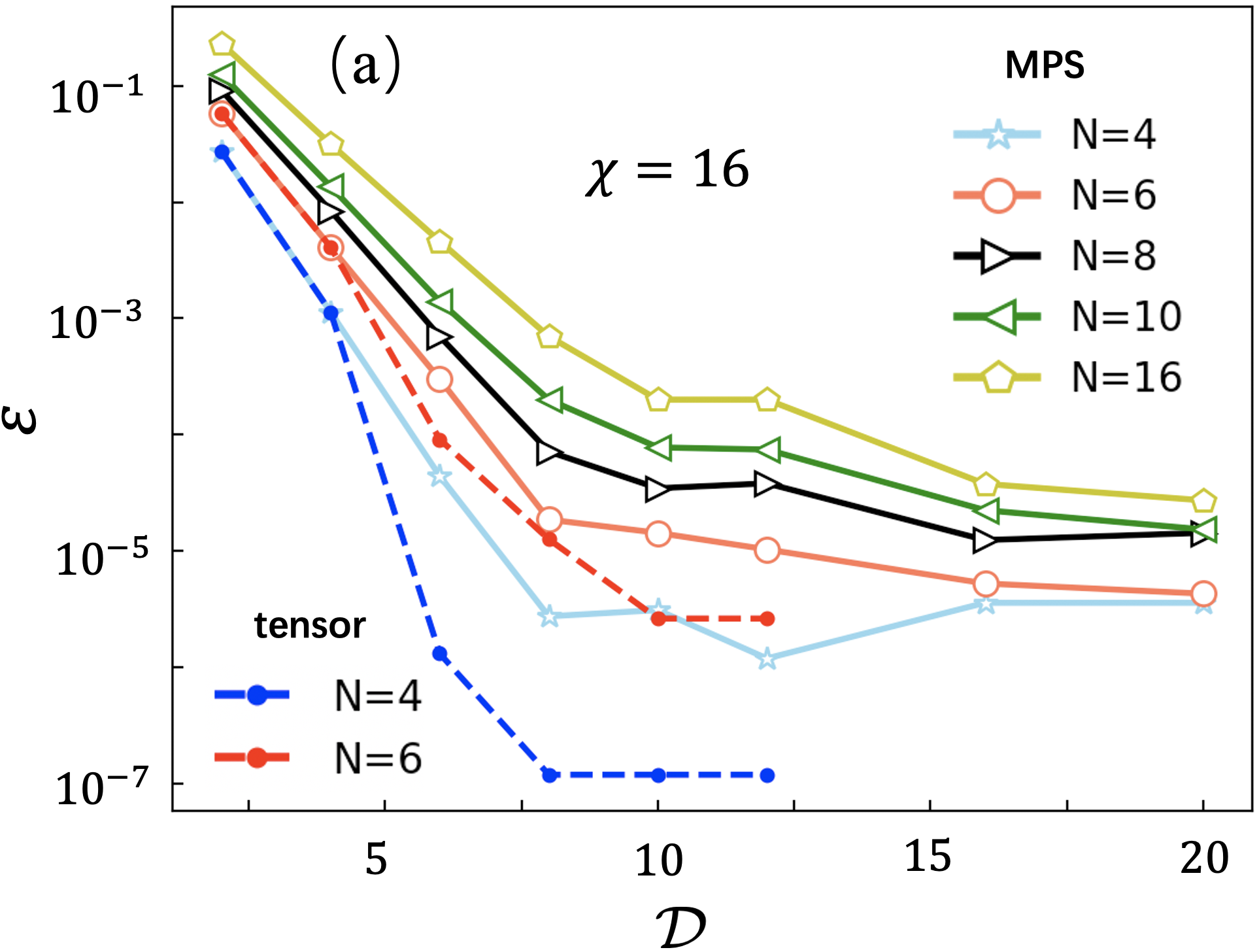}
	\includegraphics[angle=0,width=0.85\linewidth]{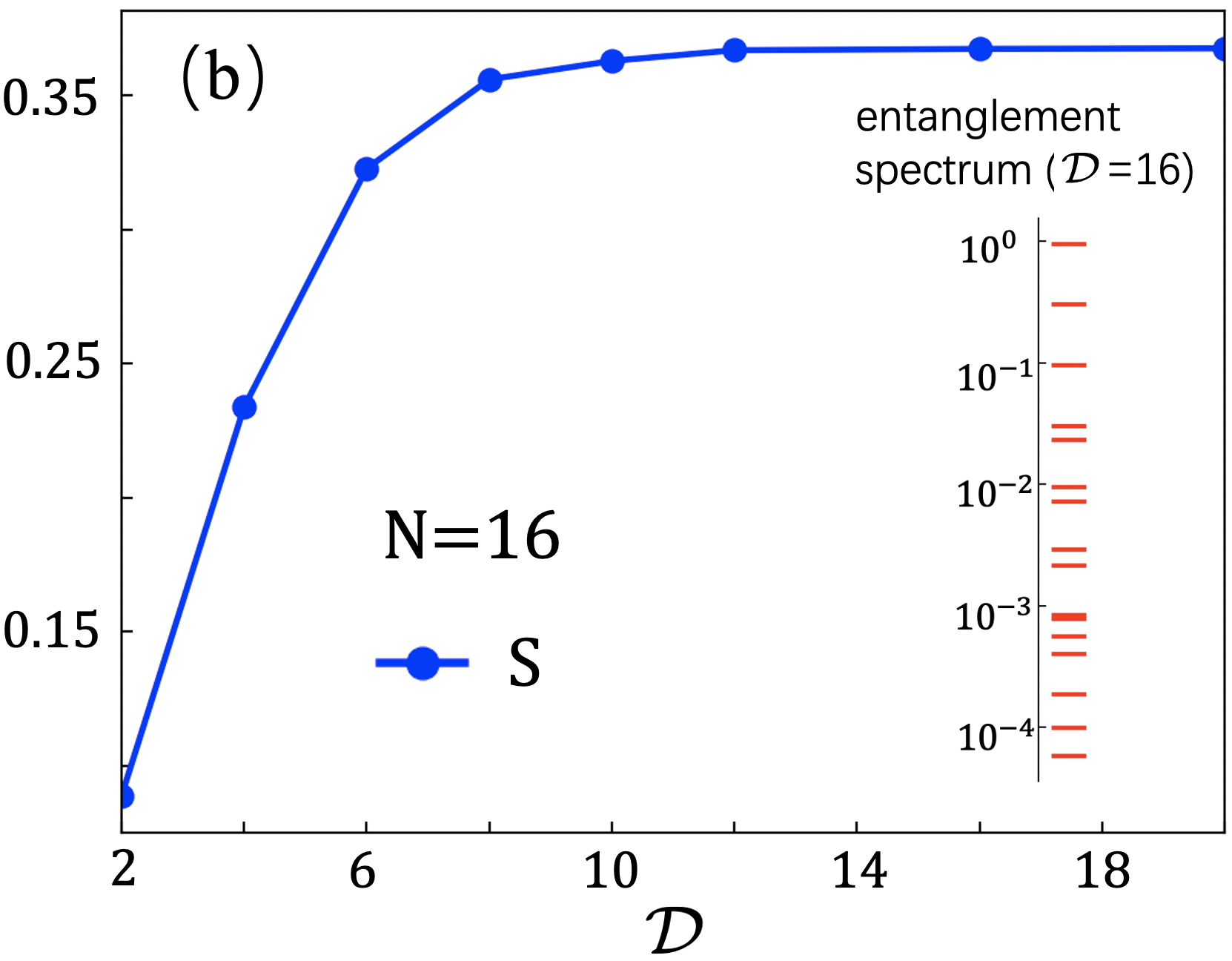}
	\caption{(Color online) (a) The error of the ground-state energy $\varepsilon$ [Eq. (\ref{eq-error})] versus the expansion order $\mathcal{D}$ for the coupled harmonic oscillators with $\gamma=-0.5$ and $\tilde{\gamma}=0$ [Eq. \ref{eq-HOH}]. We vary the number of oscillators from $N=4$ to $16$, and fix the virtual bond dimension of the MPS $\chi=16$. The results obtained by optimizing the full coefficient tensors are shown by the solid symbols with dash lines. In (b) we give the entanglement entropy $S$ [Eq. (\ref{eq-S})] versus $\mathcal{D}$ for $N=16$. The inset shows the entanglement spectrum (Schmidt numbers) measured in the middle of the MPS for $\mathcal{D}=16$.}
	\label{fig-order}
\end{figure}

Taking $\gamma=0.5$ and $N=4, 6, \cdots, 20$ as examples, the Hamiltonian in Eq. (\ref{eq-HOH}) can be exactly solved by decoupling to isolated oscillators. Fig. \ref{fig-order} demonstrates the error of the ground-state energy
\begin{eqnarray}
	\varepsilon = |E - E_{\text{exact}}|,
	\label{eq-error}
\end{eqnarray}
where $E_{\text{exact}}$ is the exact solution satisfying~\cite{coupled_2015}
\begin{eqnarray}
	E_{\text{exact}}=\frac{1}{2}\sum_{n=1}^N\sqrt{1+2\gamma cos\left(\frac{n\pi}{N+1} \right)}.
	\label{eq-exactE}
\end{eqnarray}
The hollow symbols with solid lines show the results by the function MPS method with bond dimensions $\chi=16$. For comparison, the solid symbols with dash lines show the results by directly treating the coefficients as a $\mathcal{D}^N$-dimensional tensor (as explained in Sec. \ref{sec-tensor}). For about $N<8$, the differences between the results by MPS and by tensor are small. This indicates that the errors are mainly from the finiteness of the expansion order $\mathcal{D}$ (the physical bond dimension of the MPS). The error decreases by increasing $\mathcal{D}$, and approximately converges for about $\mathcal{D}>12$. For the relatively large $\mathcal{D}$, the error by using MPS are lower than those by tensor due to the finiteness of $\chi$ in the MPS (i.e., truncation error). The differences are still slight [$\sim(O^{-5})$ or less].

Another critical advantage of our approach over the neural-network solvers is the interpretability. Below, we consider the entanglement of MPS. Thanks to orthonormal property of the functional bases, the entanglement of the MPS representing the coefficients of the wave-function shares the same quantum probabilistic interpretation of the MPS representing the quantum states of lattice models. In specific, it characterizes the ``quantum version'' of correlations between two subsystems. By ``subsystem'' in our examples, it means a subset of oscillators. 

Except for characterizing the quantum correlations among oscillators, entanglement also characterizes the truncation error of MPS induced by the finiteness of $\chi$. The entanglement entropy is defined as 
\begin{eqnarray}
	S = -2\sum_{k=0}^{\chi-1} \lambda_k^2 \ln \lambda_k,
	\label{eq-S}
\end{eqnarray}
with $\lambda_k$ the $k$-th number in the entanglement spectrum or the $k$-th Schmidt number. The upper bound of $S$ for an MPS with virtual bond dimension $\chi$ satisfies $S \sim \ln \chi$. Considering an extreme case with $S=0$, there will be only one nonzero Schmidt number. The state will be a product state $\psi(\mathbf{x}) = \prod_n \left[\sum_{s_n} C^{(n)}_{s_n} \phi_{s_n}(x_n) \right]$, and the coefficient tensor will be a rank-1 tensor satisfying $\mathbf{C} = \prod_{\otimes} \mathbf{C}^{(n)}$. For $S>0$, the truncation error in general has a same or smaller order of magnitude with the smallest Schmidt number. Obviously, we always have $S>0$ for the electronic wave-functions in order to respect the anti-commutation relations. We leave the electronic systems to our future study.

Fig. \ref{fig-order}(b) shows the entanglement entropy $S$ against $\mathcal{D}$, where the we measure $S$ in the middle of the MPS. In other words, $S$ gives the entanglement entropy between the first $N/2$ oscillators and the rest (we take $N$ to be even, without losing generality). As $\mathcal{D}$ increases, $S$ converges to about $0.36$, indicating that the ground state is not highly entangled. The inset shows the entanglement spectrum for $\mathcal{D}=16$. The smallest number in the spectrum is about $O(10^{-5})$, which is consistent with the error $\varepsilon$.

\begin{figure}[tbp]
	\centering
	\includegraphics[angle=0,width=1\linewidth]{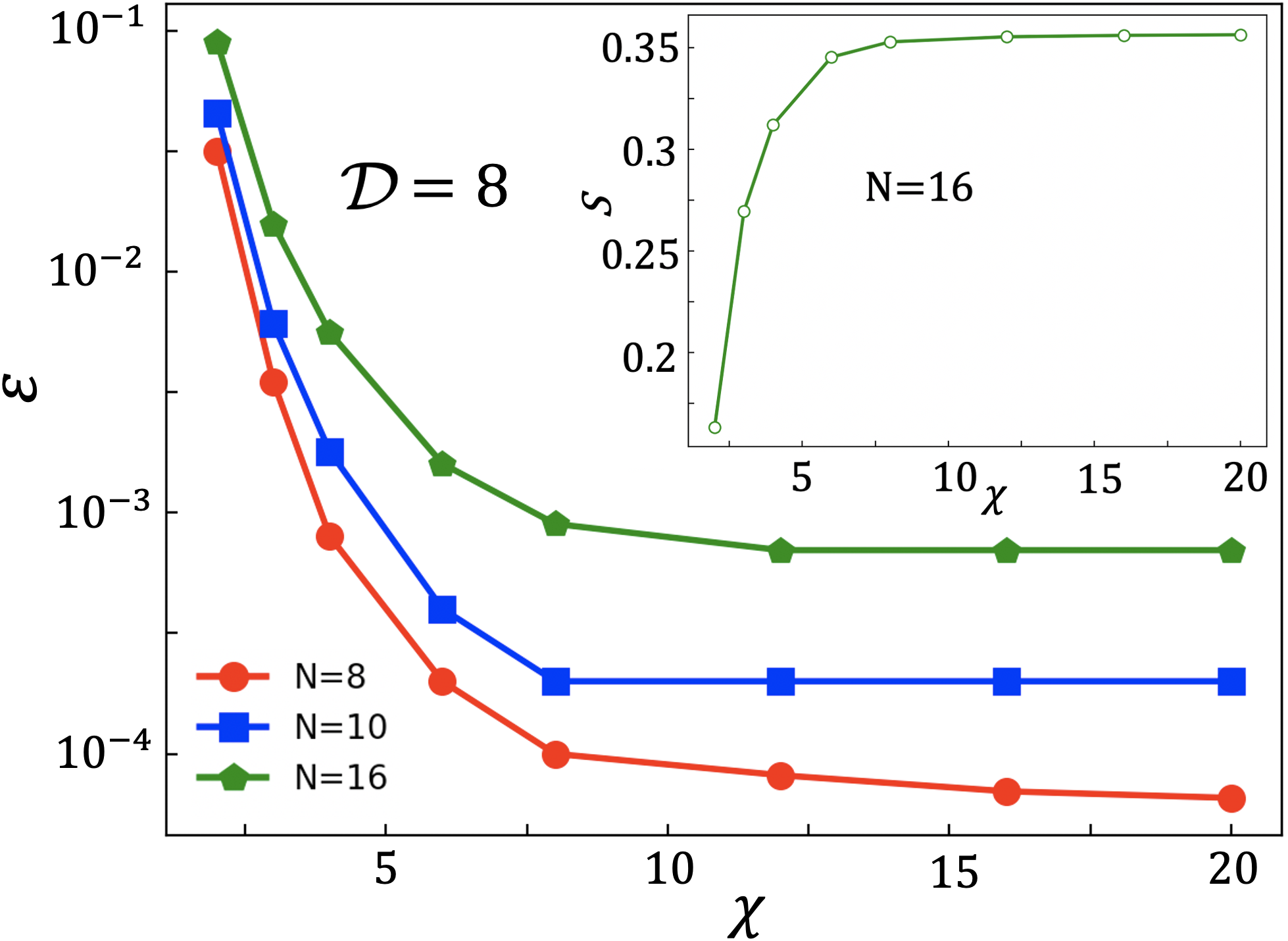}
	\caption{(Color online) The error of the ground-state energy $\varepsilon$ [Eq. (\ref{eq-error})] and entanglement entropy $S$ [inset; see Eq. (\ref{eq-S})] versus the virtual bond dimension $\chi$. We take $\mathcal{D}=8$, $\gamma=-0.5$, and $\tilde{\gamma}=0$.}
	\label{fig-chi}
\end{figure}

To further control the truncation error, Fig. \ref{fig-chi} shows that the error of energy $\varepsilon$ converges to $O(10^{-5})$ for $\chi \geq 16$ (with $\mathcal{D}=8$). When $N$ increases, $\varepsilon$ will generally increases slightly with $\chi$ remaining the same. The inset shows the entanglement entropy $S$ increases with $\chi$, meaning more entanglement will be captured (more product states are contained in the overlap) with larger $\chi$. $S$ converges to about $S\simeq 0.36$ for the $\chi\geq 16$.

\begin{figure}[tbp]
	\centering
	\includegraphics[angle=0,width=0.95\linewidth]{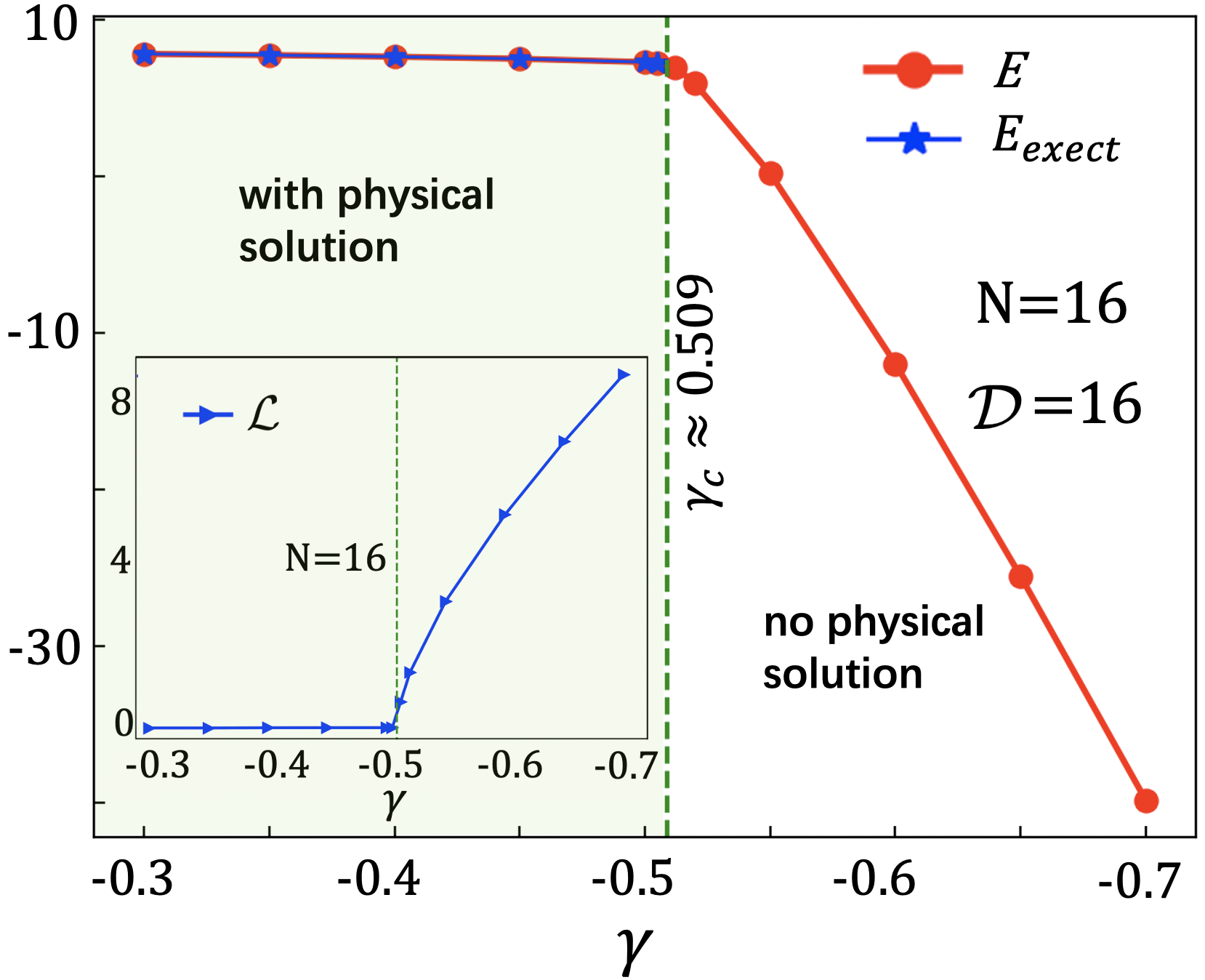}
	\caption{(Color online) The ground-state energy obtained by our method $E$ and by the exact solution $E_{\text{exact}}$ with $N=16$. For $|\gamma| > \gamma_c= \frac{1}{2} \sec \frac{\pi}{17} \simeq 0.509$, there is no real solution for the ground-state energy. The inset shows that this region can be identified by the loss $\mathcal{L}$ defined in Eq. (\ref{eq-mathcalL}), where we have $\mathcal{L} \gg 0$ for $|\gamma| <\gamma_c$. We take $N=16$ and $\mathcal{D}=16$.}
	\label{fig-phy}
\end{figure}

With different coupling strength, there is not always a ``physical'' solution. Assuming $\gamma$ to be a real number, the Hamiltonian is hermitian and the energy (an eigenvalue) should be real. From the analytical solution given by Eq. (\ref{eq-exactE}), a real solution exists for $|\gamma| < \gamma_c$ with
\begin{eqnarray}
	\gamma_c = \frac{1}{2} \sec \frac{\pi}{N+1}.
	\label{eq-gamma_cN}
\end{eqnarray}
However, the MPS still gives a converged energy even when a real ground-state energy does not exist. As shown in Fig. \ref{fig-phy}, the obtained energy matches accurately with the exact when the real energy exists. To numerically identify the region with no physical solution, we calculate the loss $\mathcal{L}$ defined in Eq. (\ref{eq-mathcalL}), which characterizes the violation of the Schr\"odinger equation. In the inset of Fig. \ref{fig-phy}, we show that $\mathcal{L}$ identifies the regions with or without a physical solution, where we have $\mathcal{L} \gg 0$ for $|\gamma| > \gamma_c $.

\begin{figure}[tbp]
	\centering
	\centering
	\includegraphics[angle=0,width=0.9\linewidth]{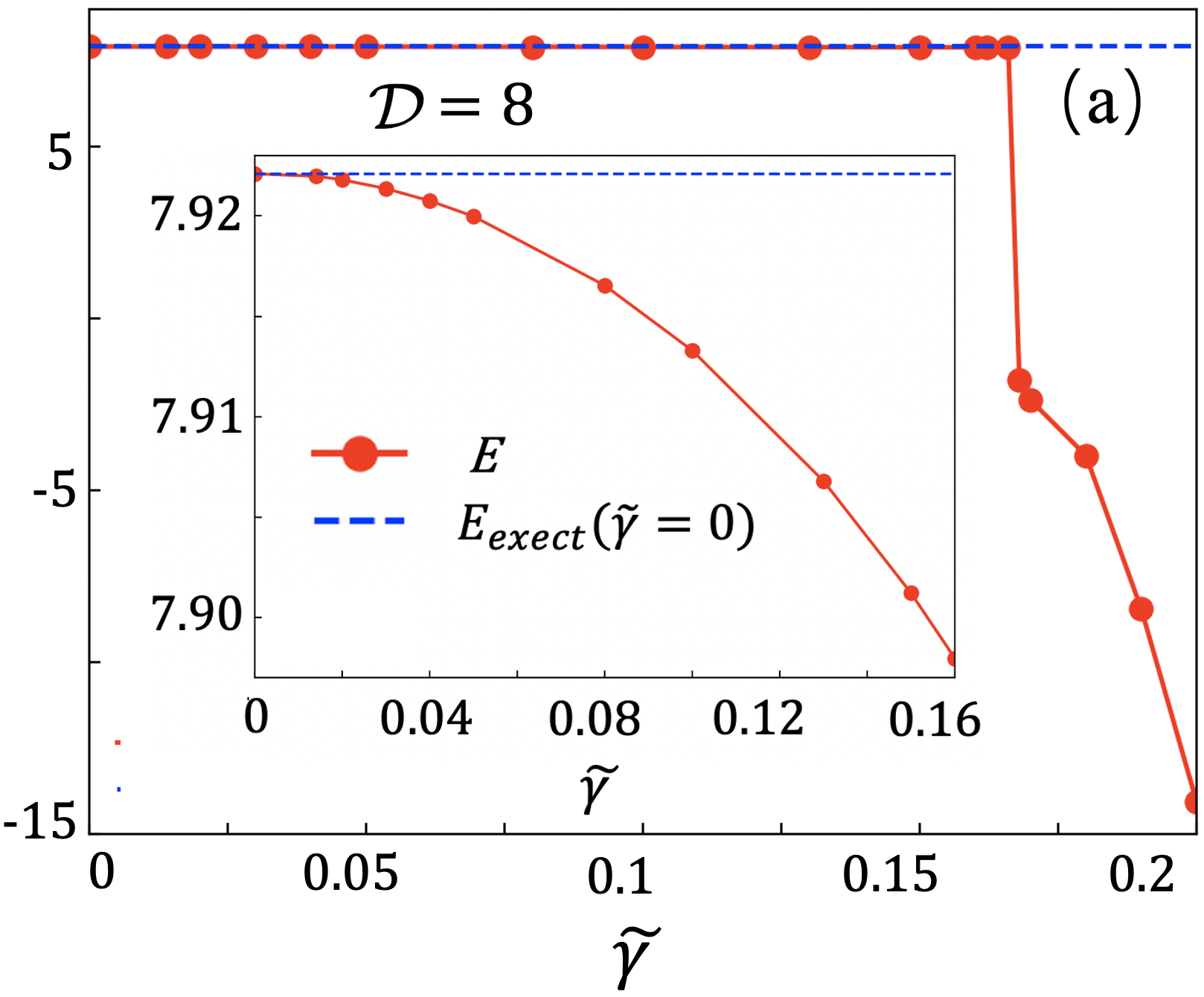}
	\includegraphics[angle=0,width=0.9\linewidth]{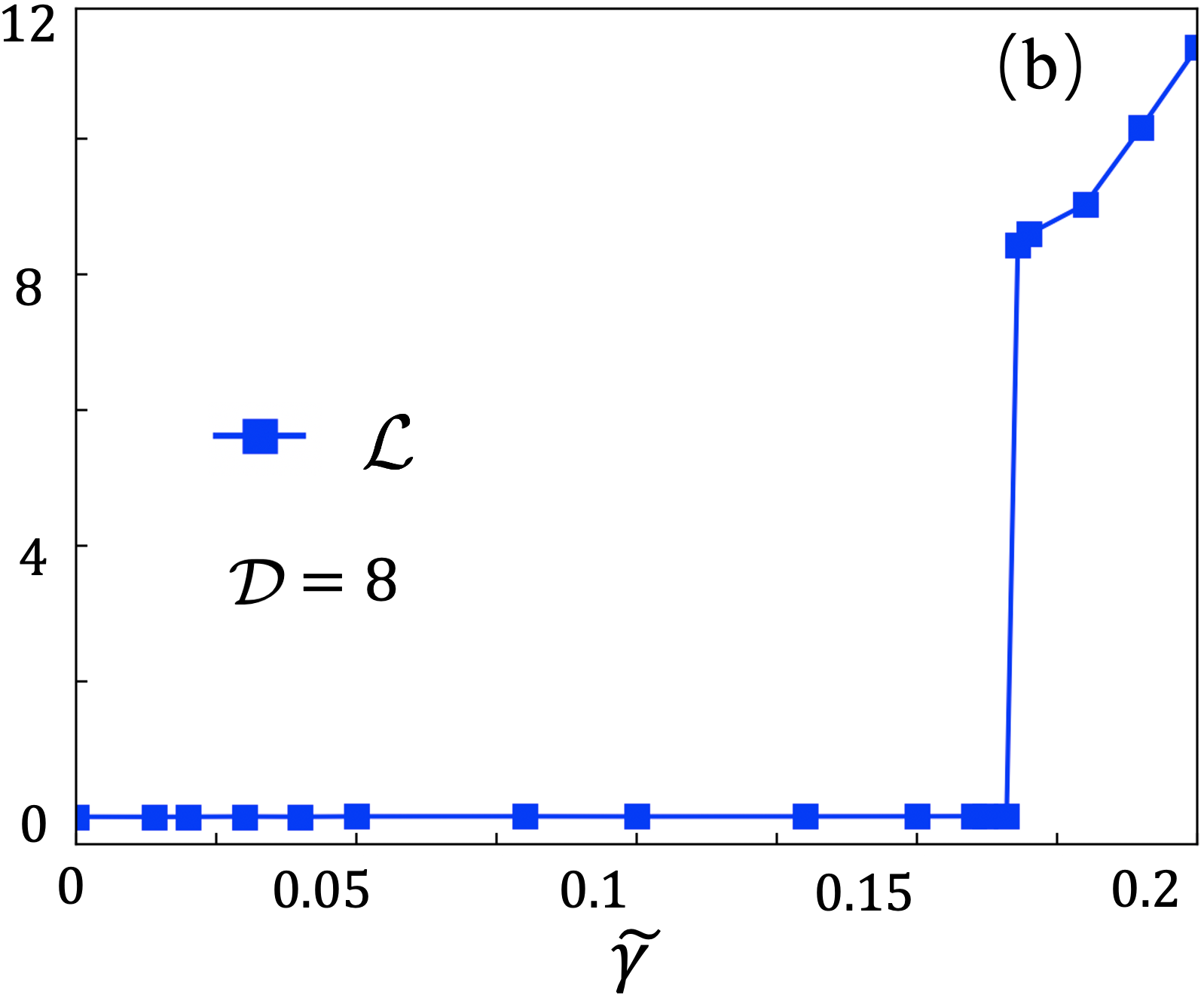}
	\caption{(Color online) (a) The ground-state energy $E$ obtained by our function MPS method versus the strength of the three-body interactions $\tilde{\gamma}$. We take $N=16$, $\mathcal{D}=8$, $\chi=16$, and $\gamma=-0.2$. The parallel dash line shows the exact energy $E_{\text{exact}}$ for $\tilde{\gamma}=0$. The inset shows $E$ for $\tilde{\gamma} < \tilde{\gamma}_c \simeq 0.168$, where there exist a real solution for the ground-state energy. In (b), we show that the loss $\mathcal{L}$ [Eq. \ref{eq-mathcalL}] suddenly becomes $\gg 1$ for $\tilde{\gamma} > \tilde{\gamma}_c$.}
	\label{fig-3body}
\end{figure}

Fig. \ref{fig-3body} (a) shows the ground-state energy $E$ obtained by our function MPS method for different three-body interaction strength $\tilde{\gamma}$ with $N=16$, $\mathcal{D}=8$, $\chi=16$, and $\gamma=-0.2$. We identify that for about $\tilde{\gamma} < \tilde{\gamma}_c \simeq 0.168$, $E$ changes smoothly with $\tilde{\gamma}$, as demonstrated in the inset. At $\tilde{\gamma} \simeq \tilde{\gamma}_c$, $E$ drastically jumps to a negative number. From Fig. \ref{fig-3body} (b), one can see that $\mathcal{L}$ suddenly becomes $\gg 1$ for $\tilde{\gamma} > \tilde{\gamma}_c$. This implies the real solution does not exists in this region.

\section{Summary and perspective}

In this work, we extend the utilization of TN to solving the many-body Schr\"odinger equation in the continuous space. Given the local functional bases, the coefficients of the wave-function are given in the form of TN, where the exponential complexity is reduced to be polynomial. The observables such as energy can be calculated simply by tensor contractions. Automatically differentiable tensors are used to form the TN. Their gradients can be obtained in a back propagation process and used to minimize the energy using gradient decent. The error of the ground-state simulation is well controlled by the entanglement. We take the TN to be MPS as an example, and apply it to the coupled harmonic oscillators with two- and three-body interactions. The existence of physical solution can be identified by the loss that characterizes the violation of the Schr\"odinger equation.

Our proposal can be readily extended to the general differential equations with many variables. The functional bases can be replaced by others, such as the Taylor series, depending on the convenience of solving the target equation. The MPS can also be generalized to other TNs such as projected entangled pair states. For electrons, the fermionic TNs~\cite{BPE09fTN, COBV10fPEPS, CJV10fPEPS, PV10fPEPS} can be used to represent the coefficients, in order to respect the anti-commutation relations. Besides entanglement, our work could build the bridge between Schr\"odinger equation and the concepts with close relevance to TN, such as symmetries~\cite{GW09TERG, SCP10PEPSsymme, SPV10TNsymme, PSGWC10symme} and quantum computation (see, e.g., ~\cite{MS08TNQcomp, AL10QCTN, FM12QCTN, DESB+18TNQsimu, HZNN+21TNQC}).

\section*{Acknowledgment} This work was supported by NSFC (Grant No. 12004266, No. 11834014, and No. 62175169), Beijing Natural Science Foundation (No. Z180013), Foundation of Beijing Education Committees (No. KM202010028013), and the key research project of Academy for Multidisciplinary Studies, Capital Normal University.

\appendix

\section{Necessary notations}
\label{append-basic}

Given an $N$-th order tensor $\mathbf{T}$, we use $T_{s_{1} s_{N} \cdots s_{N}}$ to represent a specific element. Take the following matrix (second-order tensor) $\mathbf{M}$ as an example
\begin{eqnarray}
	\mathbf{M} = \begin{bmatrix} 0 & 1 \\ 1 & 0 \end{bmatrix}.
\end{eqnarray}
We have the matrix elements as $M_{00} = M_{11} = 0$ and $M_{01} = M_{10} = 1$. Note to numbering multiple indexes or tensors, we start numbering from 1. For a given index, say $\chi$-dimensional, we take its value from $0$ to $\chi-1$. The indexes of tensors are always lower indexes. The upper ``indexes'', such as ``$(n)$'' in $\mathbf{A}^{(n)}$, are not actually indexes, but only to distinguish the symbols for different tensors. 

We use colon in the lower indexes to represent the slice of tensor, following the syntax convention of Python. For instance, $\mathbf{M}_{0, :} = [0, 1]$ is a vector that gives the zeroth raw of $\mathbf{M}$. For the $N$-th order tensor $\mathbf{T}$, we use $\mathbf{T}_{s_{1}, \cdots, s_{N-2}, :, :}$ to represent the matrix by fixing the first ($N-2$) indexes to ($s_{1}, \cdots, s_{N-2}$). The size of this matrix is $\dim(s_{N-1}) \times \dim(s_{N})$. The range of the slice can be specified. Taking a vector $\mathbf{V}$ as an example, $\mathbf{V}' = \mathbf{V}_{a:b}$ [with $a$ and $b$ two non-negative integers and $a<b<\dim(\mathbf{V})$] gives a $(b-a)$-dimensional vector, satisfying $V'_{n} = V_{n+a}$ with $n=0, \cdots, b-a-1$. Note $\mathbf{V}_{0:b}$ can be simplified to $\mathbf{V}_{:b}$, and $\mathbf{V}_{a:\dim(\mathbf{V})}$ to $\mathbf{V}_{a:}$.

\section{Addition of matrix product states}
\label{append-add}

Given two MPS's formed by the tensors $\{\mathbf{A}^{(n)}\}$ and $\{\mathbf{B}^{(n)}\}$ ($n=1, \cdots, N$), respectively, the addition of these two MPS's can be written in the MPS form. Denote the tensors of the resulting MPS as $\{\mathbf{Q}^{(n)}\}$. In general, the elements of $\mathbf{Q}^{(n)}$ are zero except for the following parts
\begin{eqnarray}
  \mathbf{Q}^{(n)}_{:\chi_{1}, :, :\chi'_{1}} = \mathbf{A}^{(n)}, \nonumber \\
  \mathbf{Q}^{(n)}_{\chi_{1}:, :, \chi'_{1}:} = \mathbf{B}^{(n)}.
\end{eqnarray}
For simplicity, we assume that the sizes of $\mathbf{A}^{(n)}$ and $\mathbf{B}^{(n)}$ for $1 \leq n \leq N-1$ are ($\chi_{1} \times \mathcal{D} \times \chi'_{1}$) and ($\chi_{2} \times \mathcal{D} \times \chi'_{2}$), respectively, considering the open boundary condition. The sizes of $\mathbf{A}^{(1)}$ and $\mathbf{B}^{(1)}$ are ($1 \times \mathcal{D} \times \chi'_{1}$) and ($1 \times \mathcal{D} \times \chi'_{2}$), respectively. The sizes of $\mathbf{A}^{(N)}$ and $\mathbf{B}^{(N)}$ are ($\chi_{1} \times \mathcal{D} \times 1$) and ($\chi_{2} \times \mathcal{D} \times 1$), respectively. Then the size of $\mathbf{Q}^{(n)}$ is [$(\chi_{1} + \chi_{2}) \times \mathcal{D} \times (\chi'_{1} + \chi'_{2})$].

If the dimensions of the left virtual bonds of $\mathbf{A}^{(n)}$ and $\mathbf{B}^{(n)}$ are both one, the above equation can be simplified to
\begin{eqnarray}
  \mathbf{Q}^{(n)}_{:, :, :\chi_{1}} = \mathbf{A}^{(n)}, \nonumber \\
  \mathbf{Q}^{(n)}_{:, :, \chi_{1}:} = \mathbf{B}^{(n)}.
\end{eqnarray}
The size of $\mathbf{Q}^{(n)}$ will be [$1 \times \mathcal{D} \times (\chi'_{1} + \chi'_{2})$] instead of  [$2 \times \mathcal{D} \times (\chi'_{1} + \chi'_{2})$]. Same simplification can be made in the case that the dimensions of the right virtual bonds of $\mathbf{A}^{(n)}$ and $\mathbf{B}^{(n)}$ are both one.

Let us now consider less general cases by assuming $\mathbf{A}^{(n)} \neq \mathbf{B}^{(n)}$ only for $n = m$, otherwise $\mathbf{A}^{(n)} = \mathbf{B}^{(n)}$. In other words, the tensors in the two MPS's are the same except for the $m$-th tensor. Then $\{\mathbf{Q}^{(n)}\}$ satisfy
\begin{eqnarray}
  &&\mathbf{Q}^{(n)} = \mathbf{A}^{(n)} = \mathbf{B}^{(n)}, \text{ for } n<m-1 \text{ or } n > m+1 \nonumber  \\
  &&\left\{ 
  	\begin{array}{rcl}
		\mathbf{Q}^{(n)}_{:, :, :\chi'_{1}} = \mathbf{A}^{(n)} & \\
		\mathbf{Q}^{(n)}_{:, :, \chi'_{1}:} = \mathbf{B}^{(n)} &
	\end{array}, \text{  for } n=m-1 \right. \nonumber \\
&&\left\{ 
  	\begin{array}{rcl}
		\mathbf{Q}^{(n)}_{:\chi_{1}, :, :} = \mathbf{A}^{(n)} & \\
		\mathbf{Q}^{(n)}_{\chi_{1}:, :, :} = \mathbf{B}^{(n)} &
	\end{array}, \text{  for } n=m+1 \right. \nonumber \\
&&\left\{ 
  	\begin{array}{rcl}
		\mathbf{Q}^{(n)}_{:\chi_{1}, :, :\chi'_{1}} = \mathbf{A}^{(n)} & \\
		\mathbf{Q}^{(n)}_{\chi_{1}:, :, \chi'_{1}:} = \mathbf{B}^{(n)} &
	\end{array}, \text{  otherwise}  \right.
\end{eqnarray}
The size of $\mathbf{Q}^{(n)}$ for $n<m-1$ or $n > m+1$ is ($\chi_{1} \times \mathcal{D} \times \chi'_{1}$), same as $\mathbf{A}^{(n)}$ or $\mathbf{B}^{(n)}$ that equal to each other in this case. The size of $\mathbf{Q}^{(n)}$ for $n=m-1$ is [$\chi_{1} \times \mathcal{D} \times (\chi'_{1} + \chi'_{2})$]. The size for $n=m+1$ is [$(\chi_{1}+\chi_{2}) \times \mathcal{D} \times \chi'_{1} $]. Otherwise, the size of $\mathbf{Q}^{(n)}$ is [$(\chi_{1} + \chi_{2}) \times \mathcal{D} \times (\chi'_{1} + \chi'_{2})$].

\section{Inner product of matrix product states}
\label{append-inner}

Given two MPS's formed by the tensors $\{\mathbf{A}^{(n)}\}$ and $\{\mathbf{B}^{(n)}\}$ ($n=1, \cdots, N$), respectively, their inner product is defined as
\begin{eqnarray}
  z = \sum_{s_1 \cdots s_{N}} \sum_{\alpha_0 \cdots \alpha_{N} \atop \alpha'_0 \cdots \alpha'_{N}}  && A_{\alpha_0 s_1\alpha_1}^{(1)} A_{\alpha_1s_2\alpha_2}^{(2)} \cdots A_{\alpha_{N-1}s_{N} \alpha_N}^{(N)} \nonumber \\ && B_{\alpha'_0 s_1\alpha'_1}^{(1)} B_{\alpha'_1s_2\alpha'_2}^{(2)} \cdots  B_{\alpha'_{N-1}s_{N}\alpha'_N}^{(N)}.
 \label{eq-MPSinner}
\end{eqnarray}

Eq. (\ref{eq-MPSinner}) can be calculated in an iterative way. We start with a matrix $\mathbf{V}$ whose size is $\dim(\alpha_1) \times \dim(\alpha'_1)$ and take $V_{0,0}=1$ (note $\dim(\alpha_1) = \dim(\alpha'_1) = 1$). Update $\mathbf{V}$ by
\begin{equation}
V_{\alpha_{n} \alpha'_{n}} \leftarrow \sum_{s_{n} \alpha_{n-1} \alpha'_{n-1}} V_{\alpha_{n-1} \alpha'_{n-1}} A_{\alpha_{n-1} s_n\alpha_n}^{(n)} B_{\alpha'_{n-1} s_n \alpha'_n}^{(n)}.
 \label{eq-MPSV}
\end{equation}
Iteratively calculate $\mathbf{V}$ by taking $n$ from 1 to $N$, and finally $\mathbf{V}$ becomes a ($1 \times 1$) matrix (i.e., a scalar) since $\dim(\alpha_N) = \dim(\alpha'_N) = 1$. We have
\begin{eqnarray}
  z = V_{0,0}.
\end{eqnarray}

An efficient way of calculating Eq. (\ref{eq-MPSV}) is to first compute $\tilde{A}_{\alpha'_{n-1} s_n\alpha_n} = \sum_{\alpha_{n-1}} V_{\alpha_{n-1} \alpha'_{n-1}} A_{\alpha_{n-1} s_n\alpha_n}^{(n)}$, and then $V_{\alpha_{n} \alpha'_{n}} = \sum_{s_{n} \alpha'_{n-1}} \tilde{A}_{\alpha'_{n-1} s_n\alpha_n} B_{\alpha'_{n-1} s_n \alpha'_n}^{(n)}$. The complexities of these two scale, respectively, as 
\begin{eqnarray}
  &&O[\dim(\alpha'_{n-1}) \dim(\alpha_{n-1}) \dim(s_{n}) \dim(\alpha_{n}) ], \\
  &&O[\dim(\alpha'_{n-1}) \dim(\alpha'_{n}) \dim(\alpha_{n}) \dim(s_{n})].
\end{eqnarray}
Thus, the complexity of calculating Eq. (\ref{eq-MPSV}) scales as 
\begin{equation}
 O\{\dim(s_{n}) \dim(\alpha'_{n-1}) \dim(\alpha_{n}) [\dim(\alpha_{n-1}) + \dim(\alpha'_{n})]\}.
\end{equation}
The above way can be used to calculate the norm of a given MPS, which equals to $\sqrt{z}$.

% \bibliography{manubib, manubibran}
\input{manu.bbl}

\end{document}